# OGSA/Globus Evaluation for Data Intensive Applications

A. Demichev[a], D. Foster[b], V. Kalyaev[a], A. Kryukov[a], M. Lamanna[b],

V. Pose[c], R. B. Da Rocha[b] and C. Wang[d]

[a]*Skobeltsyn Institute of Nuclear Physics, Moscow State University, 119992, Moscow, Russia*
[b]*CERN-IT, 1211 Geneva 23, Switzerland*
[c]*JINR LIT, Joliot-Curie 6, 141980 Dubna, Russia*
[d]*Academica Sinica, Taipei, Taiwan 11529*

We present an architecture of Globus Toolkit 3 based testbed intended for evaluation of applicability of the Open Grid Service Architecture (OGSA) for Data Intensive Applications.

## *Introduction*

It is widely accepted that future generations of Grid systems will be based on the Open Grid Service Architecture (OGSA) [1]. This architecture implies the existence of an extensive set of services, which can be combined in different ways to create various systems for distributed computing and data processing. The architecture defines standard methods for creation, naming, detection and offering service instances. In addition, OGSA suggests a platform independent integration of distributed resources on the basis of Java and XML technologies and the SOAP protocol. The development of technical specifications for OGSA is carried out in the framework of the Global Grid Forum. The Globus Toolkit® 3.0 (GT3) [2] is a first realization of the OGSA principles (the public release of 3.0 version of the Toolkit appeared on June 30, 2003).

An important area of applications of the Grid technologies is experimental High Energy Physics and in particular the LHC experimental data analysis. Currently, the activity on large-scale distributed systems for LHC preparation are carried out in the framework of the EDG [3] and LCG-1 [4] project. So far, these works were based on the preceding version of the Globus Toolkit 2 (GT2) and the European Data Grid project (EDG)[3] which does not use the OGSA.

Since the OGSA should provide enhancement of reliability, flexibility, scalability and security of distributed systems, it is important to study this approach to Grid construction and evaluate it from the point of view of potential applications in High Energy Physics. To this aim, we designed and realized a testbed (named 'Beryllium') using several PCs located at CERN and Skobeltsyn Institute of Nuclear Physics modeling a GT3 based Grid system.

## Architecture of the Globus Toolkit 3.0

The Globus Toolkit (GT3) is based on an infrastructure compliant with the OGSA, and it is an open source implementation of the Open Grid Service Infrastructure (OGSI) [5]. GT3 offers a run-time environment capable of hosting Grid services. The run-time environment mediates between the application and the underlying network OS, and transport protocol engines. A Grid service is a stateful transient service compliant to the OGSI and which exposes itself through a Web Service Description Language (WSDL) interface [6]. Currently identified sets of the toolkit services are as follows:

- GT3 Main Core Services:
    - The Admin Service is used to "ping" a hosting environment and to cleanly shut down a container.
    - The Logging Management Service allows modification of log filters and to group existing log producers into more easily manageable units at run time.
    - The Management Service provides an interface for monitoring the current status and load of a Grid service container. It also allows to activate and deactivate service instances.
- GT3 Security Services:
    - GT3 Security Services may control the access to Grid Services.
- GT3 Base Services:
    - Managed Job Service accept job from remote submission;
    - Index Service collects and distributes flux of information;

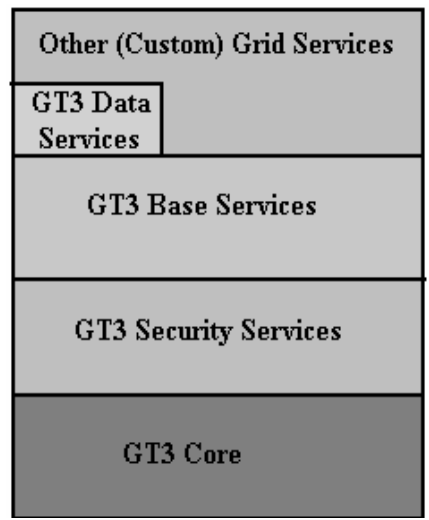

**Figure 1: GT3 Architecture**

- o Reliable File Transfer (RFT) Service is used to perform reliable file transfers;
- Other (Custom) Grid Services (high-level services that are built on top of any subset of GT3 components including the Base Services).

All these services together with the OGSI run-time environment are called the Grid Service Container. The purpose of the container is to shield the application from environment specific run-time settings, such as what database is used to persist service data. The container also controls the lifecycle of services, and the dispatching of remote requests.

OGSA proposes a factory approach to Grid Services. Instead of having one big stateless (statefull?) service shared by all users, there exists actually a corresponding central service factory which is in charge of managing transient service instances. When a client needs a new instance to be created (or destroyed), it has to talk to the factory. When a client wants to invoke some service operation, it should talk to the instance, not to the factory.

Services in GT3 have additional APIs, namely, Service Data and Notification. Service Data Element (SDE) represents structured information about the Grid Service. Now a client is able to query information on a service state (e.g., number of executed operations, type of the last operation, etc). With the help of the notification facility a Grid service may notify all Listeners (i.e., clients subscribed to the notification) about changes in its state.

### *Beryllium testbed architecture*

The primary goal of developing the testbed is to learn the potential of OGSA technology with a prototype activity. The first prototype is a simplified distributed batch system, featuring a Resource Broker (RB) and a Logging and Bookkeeping service (L&B) which are not provided by GT3, therefore completely custom, an Information Index service (II) and using GRAM built into GT3. Some of the services mimic some of the EDG component, on a much more simplified scale and with a different scope. However, the testbed architecture has some specific features that add some new functionality. One of them is a reservation of a resource for a job. Another specific property is strict separation of information flows from job and its data flows. Therefore, RB does not pass jobs and their data to resources.

A general schema of the testbed is presented in Figure 2. The system consists of the following main components:

- Resource Broker (RB);
- Information Index (II);
- Logging and Bookkeeping system (L&B);
- Computing Elements (CE);
- User Interfaces (UI).

Dotted lines represent information exchanges, while the solid lines represent substantial data transfer.

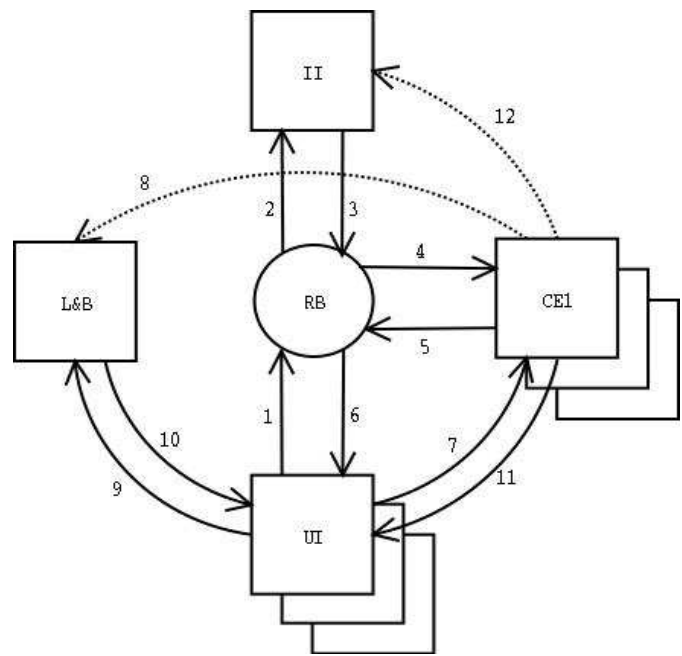

**Figure 2: General schema of Beryllium Testbed**

Functionality of the system is provided by both the high-level base services built-in the GT3 (such as, e.g., the Registry Service, MMJFS (Master Managed Job Factory Service), etc) and services specially created by us for the given testbed by using the standard methods offered by the Globus Toolkit 3.0. The arrows on Figure 2 represent steps of a cycle of job execution in the testbed.:

1. User's call to the Resource Broker for a job submission by sending job requests;
2. RB call to the Information Index for information on available resources in the system to satisfy the job requests;
3. The reply from II, RB processes the available resource list and select most suitable CE;
4. RB call to the chosen Computing Element for a reservation of its resources for the job under consideration;
5. Confirmation of the reservation from the CE's side, otherwise the CE rejects the job and the RB repeats the request to the II (point 2);
6. The resource URI and job ticket transmitted to the User Interface;
7. Automatic job submission together with the ticket to the CE (after ticket validation at the CE, the CE starts the job);
8. Notification of the Logging & Bookkeeping system about status of the job;
9. User's inquiry about job status;
10. Reply to the user on the job status;
11. Retrieving of results after the job completion;
12. Permanent monitoring of availability of resources in the system.

## *Example of Grid Services for the Beryllium testbed: CE Confirmation Service.*

In order to create a custom GT3 service, one should prepare the following:
- Service Interface. There are two options to do this:
    - to prepare Java-interface (*Name.java* ) and then to use a special GT3 tool to convert it into GWSDL file (GWSDL is an extension of WSDL);
    - to write directly GWSDL-interface (*name_port_type.gwsdl* )
- Service implementation (*NameImpl.java* )
- Deployment Descriptor (*package-config.wsdd* )
- Build-file for the Apache Ant (*build.xml* )
- In case of a service with SDE, one needs an additional input file: Service Data Description (*Name_state.xsd* )
- Client (*NameClient.java* ).

Services are deployed into the GT3 container with the help of the Apache Ant which is a Java build tool (XML-based analog of the classic *make* tool).

The CE confirmation service is responsible for confirmation of resource reservations and for ticket validation at the CE.

The service receives from the Resource Broker a call for reservation of the CE resources in the form of an incomplete (without the CE URL) ticket (Java class JobTicket). If the CE accepts the call and reserves the resources, it completes the ticket by inserting its URL and returns it to the Resource Broker. In addition, the service writes the job ID number in the special file for subsequent verification validity of the ticket coming with the job from user. If CE refuses the call, it returns "null".

Thus, as input, the service receives an incomplete ticket issued by the Resource Broker.
As an output the service produces:
- in case of accepting the call for a reservation:
    - the ticket ID which is written to a special file ("database");
    - the completed total ticket (with inserted URL of the CE) and which is returned to the RB;
- in case of rejection, the service returns "null" to the RB.

The service is realized as a Java package: gta.beryllium.ceconfirm. The interface is provided by the class Confirm.java with two methods: tktconfirm(JobTicket tkt) and getValue(), both of the JobTicket type (defined in the gta.beryllium.JobTicket).

In addition, a special script verifies tickets coming with the job from users by means of comparison with tickets, obtained from the Resource Broker. When a job with an RSL (Resource Specification Language) file containing the name of this script (written according to a special RSL-template) comes to a CE, the ticket from the RSL file is compared with the one obtained from RB and stored in the special database file. If the job ticket matches the one issued by RB, the job starts running. If the matching failed, the job is aborted and the corresponding message goes to the user.

## *Conclusion*

Until now only a preliminary version of the testbed has been realized. It provides only very basic functional interconnections of the system components and uses the simplest (simulating) algorithms for CE choice (by the Resource Broker), for acceptance/rejection of the jobs (by CEs) as well as for real resource reservation.

However already at this stage of construction and testing the system, we conclude that:

- GT3 is a very convenient environment for the creation of various specialized services which are the basic components for a distributed data processing system.
- The concept of factories of service instances has proved to be very useful. Among other advantages, this approach allows to optimize working load at a given service and to enhance its reliability. However the current implementations have some basic performance issues.
- Very important features of the services, which can be constructed with the help of the GT3 are the Service Data Elements and Notification. The former allows users and other services inquiry about current state of a given service. The latter allows to get notifications from services whenever any change in their state occurs (if a user or a service subscribed to such a notification). These features provide a convenient way to build monitoring systems.

Further information on the Beryllium testbed and related activity can be found on the Web page of GTA subproject [7].

## *Acknowledgment.*

The authors are much grateful to Les Robertson for supporting the work.
A.K. and A.D. acknowledge partially supported by CERN-INTAS grant 00-0440.